\documentclass[prl,twocolumn]{revtex4-1}
\usepackage{graphicx}
\usepackage{color}

\begin{document}

\title{Vector, Bidirector and Bloch Skyrmion Phases Induced by Structural Crystallographic Symmetry Breaking}

\author{ K. C. Erb}
\author{J. Hlinka}
\thanks{Corresponding author. Email: hlinka@fzu.cz}
\affiliation{Institute of Physics of the Czech Academy of Sciences\\%
Na Slovance 2, 182 21 Prague 8, Czech Republic}

\date{\today}

\begin{abstract}
The 212 species of structural phase transitions which break macroscopic symmetry are analyzed with respect to the occurrence of
 time-reversal invariant vector and bidirector order parameters. The possibility of discerning the orientational domain states of the low-symmetry phase by these `vectorlike' physical properties has been derived using a computer algorithm exploiting the concept of polar, axial, chiral and neutral dipoles. It is argued that for species $32>3$, $422>4$ and $622>6$,   Bogdanov-Yablonskii phenomenological theory for a ferromagnetic Bloch Skyrmions applies also to the ferroelectric Bloch Skyrmions. In these fully-ferroelectric and nonferroelastic species, the Ginzburg Landau functional allows a pseudo-Lifshitz invariant of chiral bidirector symmetry, analogous to the chiral Dzyaloshinskii-Moria term assumed in magnetic Bloch Skyrmion theory.
\end{abstract}

\pacs{11.30.Qc, 77.80.-e, 77.80.Bh, 61.50.Ah, 77.80.Dj}

\maketitle 

Symmetry lowering of the atomic and electronic spatial density distributions in crystals is a rather common phenomenon. When the point group symmetry is lowered in response to
 an isotropic influence like a temperature change, we deal with a macroscopic symmetry breaking. Such symmetry-breaking  phase transitions can be recognized by formation of orientational domains and by the appearance of new material properties. It is known that point group symmetry allows one to distinguish 212 distinct species of macroscopic symmetry reductions \cite{Hlinka2016}. Among others, attribution to a species immediately implies how many new (spontaneous) components of a given tensorial property appear in the low-symmetry phase or how many orientational domain states can be formed \cite{Hlinka2016}.

Two such properties, spontaneous polarization and spontaneous strain, play a very unique role because they are conjugated to the two most readily available anisotropic thermodynamic forces: the  electric field, and the stress tensor \cite{Aizu}. Nevertheless, species with spontaneous axial vector \cite{Hlinka2016}  and chiral bidirector \cite{Erb2018} were recently also investigated. These species describe ferroaxial \cite{Johnson2012} and chiroaxial \cite{Erb2018} phase transitions, respectively.

\begin{figure*}[ht]
    \includegraphics[width=0.7\textwidth]{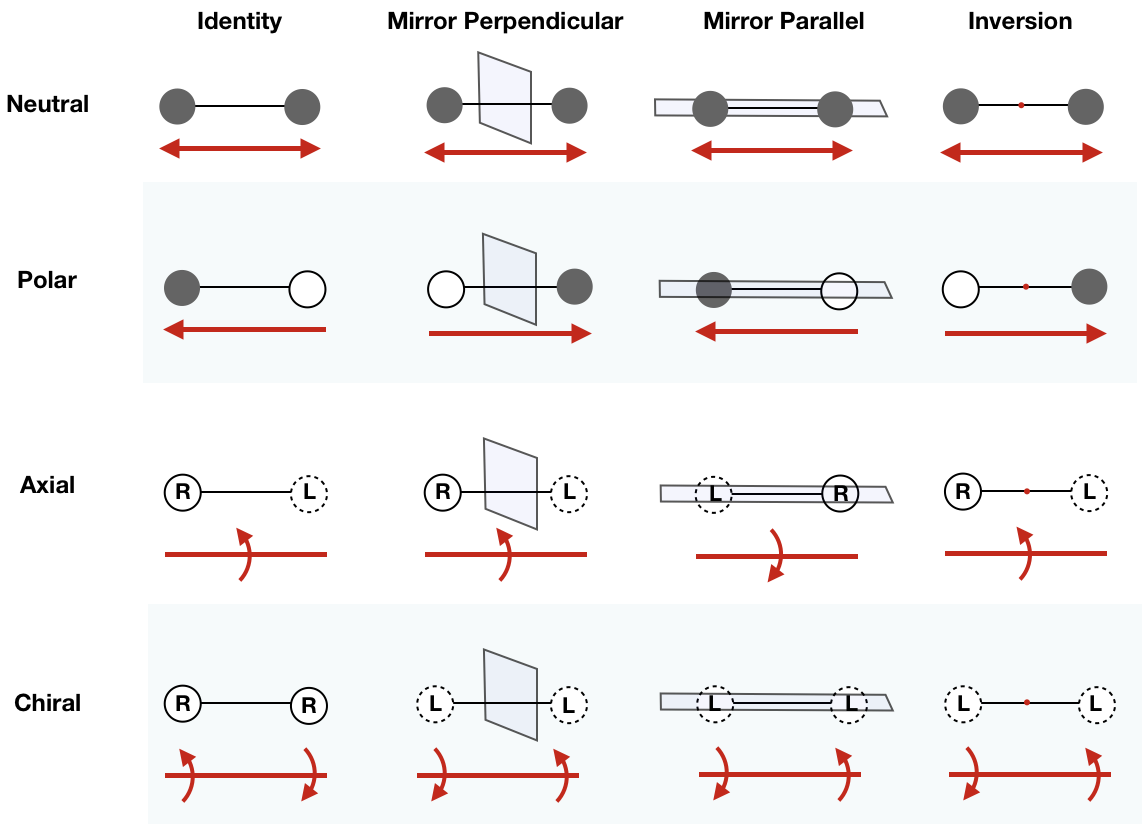}
  \caption{Transformation properties of four possible time-invariant dipoles. Each row corresponds to one of the dipoles. The first column indicates the name of the dipole property in the row. Each subsequent column indicates the effect of one selected isometry operation on these dipoles, which is either an invariance or a sign reversal. Full and empty circles stand for points with an opposite scalar property  while circles with R or L inside stand for points with an opposite pseudoscalar property (right-handed and left-handed points). Otherwise, the points do not differ in any other internal property. }
  \label{fig:dipoles}
\end{figure*}

From the point of view of rotational symmetry, the polar vector, axial vector, chiral bidirector, and neutral bidirector cover all possible types of
properties that possess only a single axis, modulus, and a binary sign \cite{Hlinka2014} (see Fig.\,1).
Therefore, occurrence of all these four  vectorlike spontaneous properties is needed for any complete analysis on the axial properties of species.

In the present work, a computer algorithm has been designed to detect whether orientational domain states can be fully or partially distinguished by a given vectorlike property. For this purpose, we generalized the concept of electric dipole.
Among many multiferroic cases, where several vectorlike spontaneous properties are encountered simultaneously, phases with ferroelectric Bloch Skyrmion symmetry deserve special attention, where polar, axial, and chiral ``dipole" can appear at the same axis \cite{Hlinka2014}. We extended the argument of Bogdanov and Yablonskii \cite{Bogdanov1989} and proposed that ferroelectric Skyrmion phases can be found in crystals undergoing ferroelectric phase transitions belonging to species 31, 70 and 100.

Before presenting our results, let us note that a macroscopic crystal symmetry  is defined by a crystallographic point group - set of all  $O(3)$ isometries (proper and improper rotations around a fixed point in 3D space) that preserve all macroscopic material properties of the crystal. If there is an isometry in $O(3)$ that transforms one such symmetry group to another one, we consider the two
 groups as crystallographically equivalent ones. All symmetry groups linked directly or consequentially by such equivalence relation form a 
 crystallographic
 class. Any crystal with 3D translational periodicity belongs to one of the 32 crystallographic classes \cite{Newnham2005, Janovec2017}.

Macroscopic symmetry {\em reduction} is defined by group-subgroup relation $G>F$ between point groups $F,G$ of a crystal in its higher and lower symmetry phase, respectively. If there is an isometry in  $O(3)$ that transforms the group-subgroup pair $G>F$ into another pair $G'>F'$, the two pairs are crystallographically equivalent. All symmetry pairs linked by a chain of such equivalence relations constitute a macroscopic symmetry breaking species, or simply a species. There are 212 such species \cite{Hlinka2016,Aizu}.

The existence of lost-symmetry elements generates several distinct but energetically equivalent domain states of phase $F$. Number $n$ of these orientational domain states is given by the quotient of the high symmetry group's order and the lower symmetry group's order \cite{Janovec}.
 In general, point group symmetry allows one to answer the following questions:
(i) is a property allowed by the symmetry $F$ of the crystal, and is there any restriction on the property orientation?
(ii) is the $G>F$  symmetry lowering  removing some symmetry restriction on  this property?
(iii) does the sign or orientation of this property allow one to distinguish all or some orientational domain states?

\begin{table}
\begin{tabular}{c|l|l}
Symbol         & Limiting group & Type of the property
 \\ \hline
N      & $\infty/mm$    & Neutral bidirector \\
P      & $\infty m$     & Polar vector      \\
G      & $\infty/m$     & Axial vector      \\
C      & $\infty 2$     & Chiral bidirector \\
\end{tabular}
\caption{Limiting groups defining the symmetry of four time-reversal invariant vectorlike quantities.}
\label{tab:notation-table}
\end{table}

The  question (i) is formally solved by the Neumann principle, which states that a property is allowed if the crystal class  is a subgroup of the symmetry of the property \cite{Newnham2005}. In this paper, we deal with the 4 types of vectorlike properties labeled as $\bf{N}$, $\bf{P}$, $\bf{G}$, and $\bf{C}$. Their symmetry is given by the corresponding Curie groups listed in Table\,I. A simple object with polar vector symmetry is electric dipole, a pair of points of with opposite charges. Interestingly, we note that a similarly simple
 object with the neutral bidirector symmetry $\bf{N}$ is a pair of points of with equal charges. Likewise, a pair of points of with pseudoscalar properties of opposite signs has symmetry of the axial vector $\bf{G}$, and a pair of points of with an identical pseudoscalar property has a symmetry of chiral bidirector $\bf{C}$. Direct inspection of symmetry invariance of these four ``dipoles" is displayed in Fig.\,1.  The question (i) can be thus reformulated as whether and how such a corresponding ``dipole" can be placed in the crystal  without breaking its point group symmetry.

 In practice, we have designed a computer algorithm that allows one to place a generalized dipole of any above type in the center of crystallographic reference frame and to check whether all symmetry operations of a given point group leave such a dipole intact or not. By going through the general and all special positions with respect to the symmetry elements in the group, possible location of dipoles are found.
 In general, polar and axial vectors, if allowed, are restricted to a crystallographic axis or within a crystallographic plane, or not restricted at all. In contrast, bidirectors, if allowed, could be restricted to an axis, or to a triplet of orthogonal axes, or to a plane, or a plane and the perpendicular axis, or not restricted at all. For example, Fig.\,2 summarizes restrictions of the four vectorlike properties within pointgroups 222 and 2.

\begin{figure}[ht]
    \includegraphics[width=0.35\textwidth]{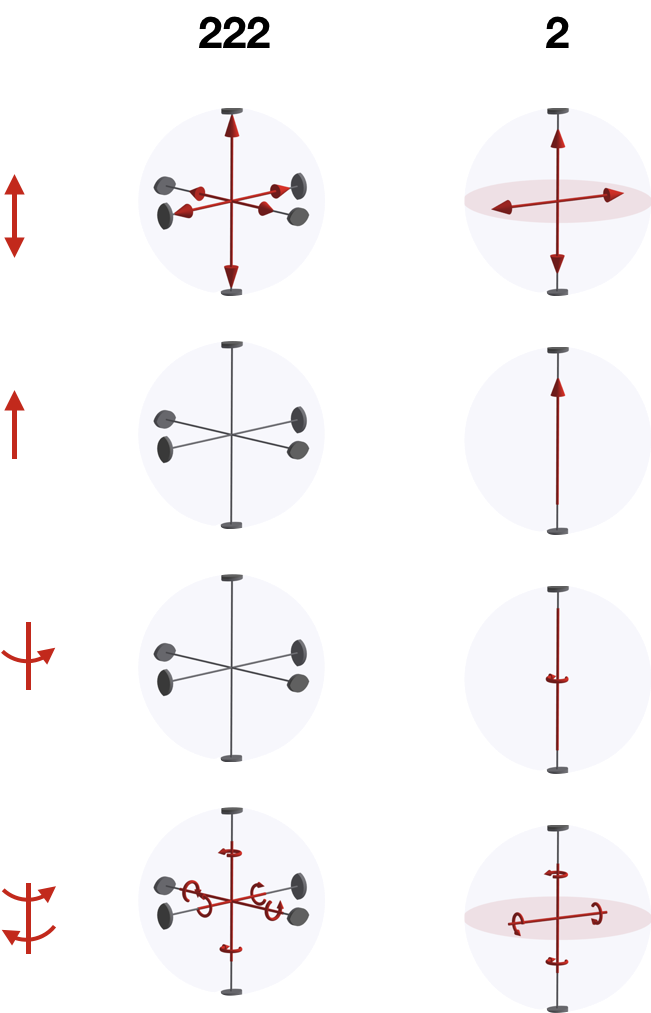}
  \caption{Symmetry-allowed orientation of vectorlike  properties in crystals of 222 and 2 symmetry. Columns correspond to point groups, rows to  $\bf{N}$, $\bf{P}$, $\bf{G}$, and $\bf{C}$-type quantities.
  The symmetry unrelated bidirectors  are of different magnitude. Under the $222>2$ symmetry lowering,  polar and axial vectors appear along the vertical diad, while the bidirectors perpendicular to it are released to have an arbitrary orientation within the indicated plane). }
  \label{fig:rep_222_2}
\end{figure}

In order to answer the question (ii), another algorithm was designed to find a group-subgroup representative for each species. Then, the released symmetry constraints on vectorlike properties were obtained by comparison of the symmetry constraints in the group-subgroup pair. For example, the species $222>2$ gives new degrees of freedom for any of the four
vectorlike property (see Fig.\,2.)

 Finally, we have investigated whether the selected property allows to distinguish  all or some of the orientational domain states in a given species. Up to three independent generalized dipoles of a given type $X$ were placed in the center of crystallographic reference frame in the most general way compatible with our representative low-symmetry group of a given $G>F$ species. This gives unique description of the domain state in terms of given type of dipoles even in case of bidirectors.
 When isometries of the high symmetry group were applied, either the same or a different  dipole configuration is obtained. After going trough all symmetry elements of $G$, number of all distinct dipole configurations $n_{X}$ was counted and compared with the number of all orientational domain states $n$.

The numbers of ferroelectric, ferroaxial and ferrochiral domains match the earlier obtained scores \cite{Hlinka2016,Erb2018}.
Moreover, we noted that the gained neutral bidirector freedom  corresponds exactly to the degree of released symmetry constraints on the components of the  second-order symmetric polar tensor. Thus, the possibility to distinguish orientational domains states fully or partially by $\bf N$ is actually equivalent to full and partial ferroelasticity.

 The results are summarized in Fig.\,3; full distinction of orientational domain states by the property ($n_{X}=n$) is marked by the full circle, partial distinction ($1<n_{X}<n$) is marked by the half-filled circle. Species which do not release any additional degree of freedom for the property $X$  are marked by an open circle. This last situation happens either when $n_{X}=1$ or when the property is not compatible with the symmetry of  $F$ at all ($n_{X}=0$). Differentiation between the last two options is quite trivial, so we use a common label there.

\begin{figure*}[ht]
\includegraphics[width=1\textwidth]{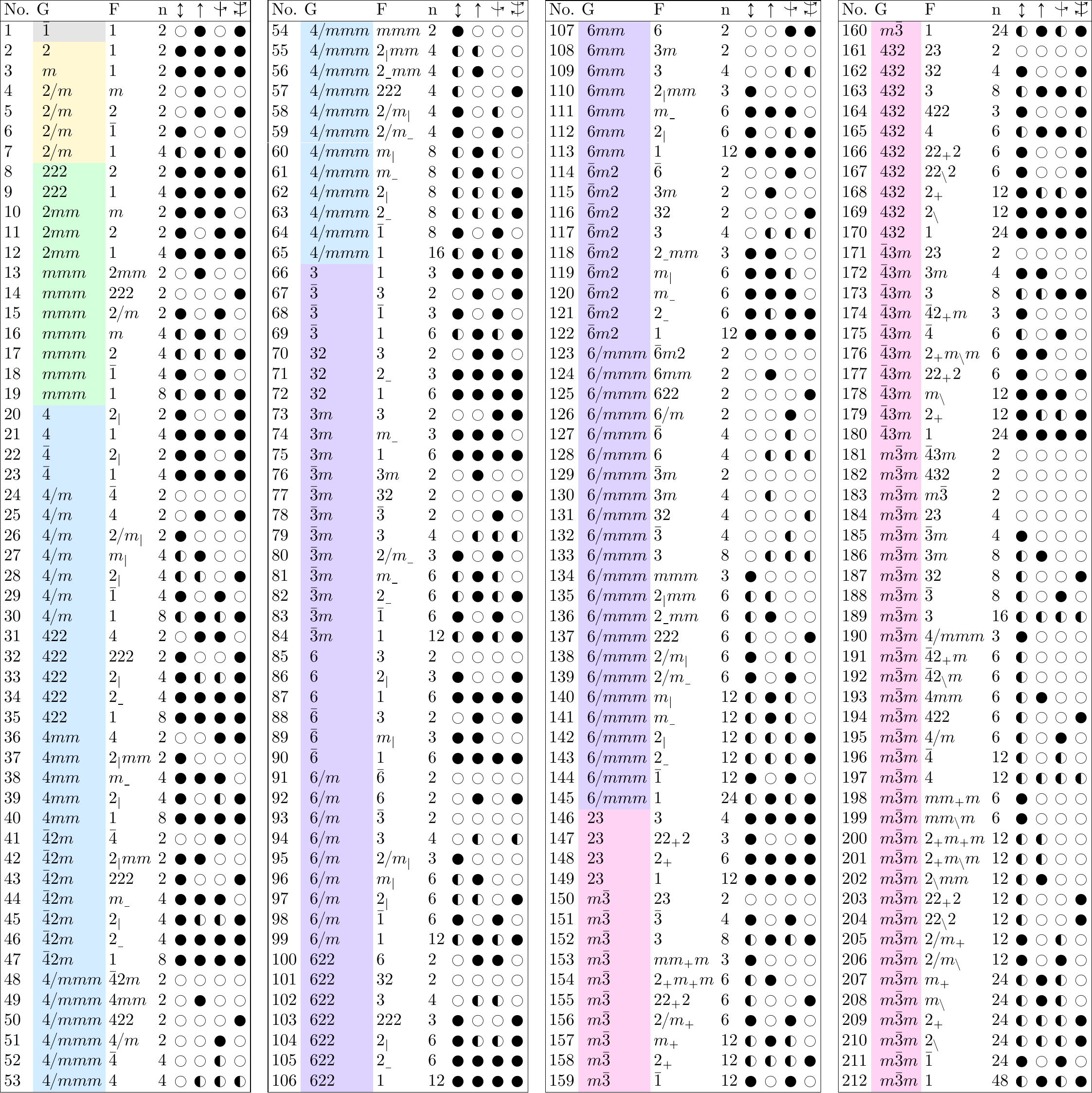}
\caption{Table summarizing possibility to distinguish orientational domains states within individual macroscopic symmetry breaking species $G>F$ fully (filled circle), partially (half-filled circle), or not at all (open circle) by each the four possible vectorlike properties separately. The number of all orientational domain states is given in the column marked $n$. See the explanations in the main text.}
\label{tab:non-magnetic}
\end{figure*}

 Finally, let us turn back to Skyrmions. Originally, the existence of magnetic analogues of Abrikosov vortexes was predicted by inspecting solutions of the Ginzburg-Landau potential for a ferromagnet with free energy density in the form \cite{Bogdanov1989}
\begin{equation}
\frac{1}{2} \alpha(\nabla {\bf M})^2 +\frac{1}{2}\beta M_{z}^2 - M_{z} H + \gamma w(\bf{M})~,
\end{equation}
where $\alpha$ is the exchange interaction prefactor,  $\beta>0$ defines the easy axis anisotropy, $H$ is the magnetic field along the axis and $\gamma$ is the strength of the Dzyaloshinskii - Moriya interaction (DMI)\cite{Dzyaloshinskii57, Moriya}.
The axisymmetric Bloch Skyrmion solution requires that the leading term of DMI is in the form \cite{Bogdanov1989}
\begin{equation}
w = M_z\frac{\partial M_x}{\partial y} -  M_x\frac{\partial M_z}{\partial y} - M_z\frac{\partial M_y}{\partial x} +  M_y\frac{\partial M_z}{\partial x}~.
\end{equation}
This expression does not depend on the rotation of the coordinates around the $z$-axis \cite{Bogdanov1989}. Actually,
direct check proves that it has the full chiral bidirector symmetry. Therefore, the
 paramagnetic phase allows for chiral bidirector property along the $z$-axis. This implies that Bloch Skyrmion model of eqs.\,(1-2) holds when the paramagnetic phase belongs to class 32, 422, 622, 3, 4, or 6, in agreement with earlier results \cite{Dzyaloshinskii, Bogdanov1989}.

And how about the ferroelectric polarization, can it also form a similar axisymmetric Bloch Skyrmion soliton?
When eqs.\,(1-2) are written for
for electric polarization and electric field, the eq.\,(1) describes a proper uniaxial ferroelectric.
 The pseudo-Lifshitz term in eq.\,(2) transforms as a chiral bidirector even when magnetization is replaced by electric polarization.
  Therefore, the symmetry limitations on the paraelectric phase are the same as before on the paramagnetic phase. Since proper ferroelectricity implies full ferroelectricity \cite{Janovec}, by inspection of Fig.\,3, it can easily be found that there are only three species that correspond to the above energy functional form: species 31, 70, and 100. Thus, thermodynamically stable ferroelectric Skyrmion phases are possible and the present analysis indicates where to look for them.

At the same time,  this Bogdanov-Yablonskii scenario of eqs.\,(1-2) is not the only possibility for the formation of ferroelectric Bloch Skyrmion textures. In principle, the primary order parameter which induces the Skyrmion phase may not be polarization. Also, it might be sufficient if the chiral DMI term emerges in the low-temperature phase only.
For example, recent observations revealed  homochiral  Skyrmion bubble nanodomain textures in a material with achiral paraelectric phases \cite{Das2019}.
 In this case, the Skyrmion symmetry breaking could be possibly treated as species 36 ($4mm>4$) or 53 ($4/mmm>4$).
 Unlike in the Bogdanov-Yablonskii scenario, here  enantiomorphic partner textures may form, with precisely the same energy and polarization but with inversed sign of Skyrmion density. Whether this is an essential advantage or disadvantage is yet to be found.


In summary, the concept of generalized dipoles allowed us to count domain states distinguishable by vectorlike properties and summarize possible property combinations. Analysis promises existence of ferroelectric Bloch Skyrmion phases. Moreover, the point-group symmetry breaking applies  to other areas of physics as well and the idea of generalized dipoles can be directly extended to multipoles and time-odd properties \cite{Tanaka2008, Lovesey2011, Cheong2018, Hayami2018, Lazzeretti2017, Lazzeretti2018, Lazzeretti2019}.

This work was supported by the Czech Science Foundation (projects no. 17-11494J and 19-28594X).

\end{document}